\documentclass[aps,prd,twocolumn,showpacs,preprintnumbers,amsmath,amssymb]{revtex4}
\usepackage{graphicx}
\usepackage{dcolumn}
\usepackage{bm}
\usepackage{epsfig}
\begin{document}
\title{Test of constancy of speed of light with rotating cryogenic optical resonators}
\author{P. Antonini, M. Okhapkin, E. G\"okl\"u\footnote{present address ZARM, University of Bremen, 28359 Bremen, Germany}, S. Schiller}
\affiliation{Institut f{\"u}r Experimentalphysik, Heinrich-Heine-Universit{\"a}t D{\"u}sseldorf, 40225 D\"usseldorf, Germany}

\vskip -.5in
\date{to appear in Phys. Rev. A, 2005}
\begin{abstract}
A test of Lorentz invariance for electromagnetic waves was
performed by comparing the  resonance frequencies of two optical resonators as a function of
orientation in space. In terms of the Robertson-Mansouri-Sexl theory, we obtain 
$\beta-\delta-1/2=(+0.5\pm 3\pm0.7)\cdot10^{-10}$, a ten-fold improvement compared to the previous best results. 
We also set a first upper limit for a so
far unknown parameter of the Standard Model Extension test theory,
$|(\tilde{\kappa}_{e-})^{ZZ}| < 2\cdot 10^{-14}$.
\end{abstract}
\pacs{03.30.+p, 12.60.-i, 07.60.-j}
\maketitle

\section{Introduction}

The principle of Local Lorentz Invariance (LLI), stating the independence of physical laws from the state of motion of inertial laboratories, 
is one of the most fundamental facts about our physical world. 
The first precision measurement, demonstrating LLI for light propagation, was performed by Michelson and Morley in 1887 and 
its result was an essential experimental foundation for the advent of Relativity.  LLI is incorporated as
a fundamental symmetry into the accepted theories of the fundamental forces, General Relativity \cite{Will} and the Standard Model.
 Numerous 
experiments have tested LLI with respect to matter and to the electromagnetic field and have upheld its validity until to date 
\cite{KosteleckyProceedings}. For electromagnetic waves  the isotropy of space has so far been verified a the level of a few parts in $10^{15}$ \cite{bri79,wol03,Lipa2003,mue03b,Wolf04}.

New generations of tests of LLI and of other fundamental symmetries (Weak Equivalence Principle, Local Position Invariance, CPT Symmetry) are seen as one important approach in the quest for a deeper
understanding of the forces of nature \cite{KosteleckyProceedings}. They might provide useful input for the development of a theory able to describe gravity at the quantum level.
In these theories, violations of fundamental symmetries are being considered.
Thus, the  theoretical models call for improved experiments to either validate LLI at
much higher precision levels, or to uncover its limits of validity.

Violations of LLI can be interpreted using so-called test theories. A kinematic test theory commonly applied is that by
 Robertson, Mansouri and Sexl (RMS) \cite{Mansouri:1977}. 
Here, light propagation is described relative 
to a preferred frame (``ether frame'') $\Sigma$ in which there is no preferred direction and thus the speed of light 
${c_0}$ is constant.
Usually the frame in which the cosmic microwave background is isotropic is assumed to be this frame. 
Lorentz transformations between a laboratory frame 
S and $\Sigma$ are replaced by general linear transformations which depend 
on the velocity $\vec{v}$ of the lab 
frame with respect to $\Sigma$ and on three phenomenological parameters $\alpha$, $\beta$, and $\delta$. 
These reduce to $\alpha = -\frac{1}{2}$, $\beta = \frac{1}{2}$, $\delta = 0$ if LLI is valid. 
In the moving frame S, the speed of light can be expressed to lowest order in $|\vec{v}|/c_0$ as
${c(\phi,v)}/{c_0}=1+({v^2}/{c^{2}_0})\left(\beta - \alpha -1+
                          \left(\frac{\small 1}{\small 2}-\beta+\delta\right)\sin^2{\phi}\right) ,$
where $\phi$ is the angle of the direction of light propagation relative to the velocity $\vec{v}$ of the laboratory. 
Thus, violations of the constancy of the speed of light are described by 
two nonzero parameter combinations.

Recently, a comprehensive dynamical test theory of LLI violation has been developed,
the Standard Model Extension (SME) \cite{Kostelecky:2002hh}. It is based on the Lagrangian of the Standard
Model, extended by terms that violate LLI and CPT. These terms describe possible violations in
the behavior of both matter and fields, and contain a large number of unknown
parameters, whose values can in principle be determined by an appropriately large set
of (partially independent) experiments. In particular, the
extended Lagrangian of the electromagnetic field is given by
$\mathcal{L}=-\frac{1}{4}F_{\mu\nu}F^{\mu\nu}
-\frac{1}{4}(k_F)_{\kappa\lambda\mu\nu}F^{\kappa\lambda}F^{\mu\nu}.$
The dimensionless tensor $(k_F)_{\kappa\lambda\mu\nu}$ describes violations of LLI
and has 19 independent coefficients. Its values are dependent on the frame of reference;
a frame in which the Sun is stationary is chosen for practical reasons.
Of these coefficients, 10 describe polarization-dependent effects. These can be restricted to
the level of $10^{-32}$ using astronomical observations of the polarization of distant light sources
\cite{Kostelecky:2001mb}.

Of the remaining 9 coefficients, one ($\tilde{\kappa}_{tr}$) describes an asymmetry of the one-way
speed of electromagnetic waves \cite{Tobar05}, while the others describe different aspects of violations of constancy, {\it i.e.} a  dependence  on propagation
direction and on the speed of the laboratory frame of reference.  The corresponding eight coefficients can
be arranged in two traceless 3$\times$3 matrices:
the antisymmetric $(\tilde{\kappa}_{o+})^{ij}$ that describes violation of boost invariance
and therefore enters observable quantities  weighted by the ratio $\beta_\oplus\simeq10^{-4}$
of Earth's orbital velocity and the speed of light, and
the symmetric $(\tilde{\kappa}_{e-})^{ij}$ quantifying anisotropy of $c$.

These eight coefficients can in principle be determined by measuring the time dependence of the resonance frequency of an
electromagnetic resonator (assuming that particles satisfy LLI). If the electromagnetic resonator is stationary, 
seven coefficients can be determined
by taking advantage of the rotation and orbital motion of Earth. Three experiments, using
ultra-stable cryogenic resonators for microwaves and optical waves, have
been performed recently along this line. Lipa {\it et al.} \cite{Lipa2003} took data for $\sim$100 days and could 
constrain four coefficients of $\tilde{\kappa}_{e-}$ and four linear combinations of three coefficients of $\tilde{\kappa}_{o+}$.
M\"uller {\it et al.} \cite{mue03b} performed an experiment where the measurement duration was
extended to a sufficient duration (over 1 year) that the measurement of the 7 coefficients
was achieved. Wolf et al. \cite{Wolf04} extended the measurement time further and improved significantly on the limits.

The eighth coefficient, $(\tilde{\kappa}_{e-})^{ZZ}$, has so far not been determined.  

\vskip .2in
\section{Method}

The resonator frequency measurement involved in a test of LLI requires a frequency
reference with a frequency instability comparable to the desired measurement accuracy \cite{wol03,Wolf04}.
An alternative approach is the use of two cavities oriented at right angles \cite{Lipa2003,mue03b}. 
One then measures the  difference between the two cavity frequencies, $\nu_1(t)-\nu_2(t)$. 
For a two-cavity configuration, the beat frequency modulation can be expressed as
\begin{equation}
\frac{\delta(\nu_1(t)-\nu_2(t))}{\nu}=
2 B(t) \sin{2\theta(t)} +2 C(t) \cos{2\theta(t)},\label{eq1}
\end{equation}
where $\nu_1\approx\nu_2\approx\nu$ ($2.8\cdot 10^{14}$ Hz) is the average frequency and $\theta(t)$ it the angle between 
one cavity's axis relative to the south direction. Each amplitude 2$B(t)$ and 2$C(t)$ is a linear combination of eight SME
coefficients weighted by time-harmonic factors. The amplitude $B(t)$ contains frequency
components at
0, $\omega_{\oplus}$,
$2\omega_{\oplus}$, $\omega_{\oplus}\pm\Omega_{\oplus}$ and
$2\omega_{\oplus}\pm\Omega_{\oplus}$,
while $C(t)$ contains in addition one component at the frequency $\Omega_{\oplus}$. Here $\omega_{\oplus}$ is
Earth's sidereal angular frequency and $\Omega_{\oplus}$ is Earth's orbital frequency.
The determination of the individual $\tilde{\kappa}_{o+}$ coefficients requires the ability to 
resolve the contribution of Earth's orbital motion in order to discriminate between frequency 
coefficients differing by $\Omega_{\oplus}$. Thus a measurement extending over of at least 1 year is necessary.

The contribution of $(\tilde{\kappa}_{e-})^{ZZ}$ to the beat frequency signal is
\begin{equation}
\frac{\delta(\nu_1(t)-\nu_2(t))}{\nu}=\frac{3}{4}(\tilde{\kappa}_{e-})^{ZZ}\sin^{2}{\chi}
\cos{2\theta(t)}+...\label{eq2},
\end{equation}
where $\chi$ is the colatitude of the laboratory. 

It is evident from eq.(\ref{eq2}) that $(\tilde{\kappa}_{e-})^{ZZ}$  
coefficient is only measurable by active rotation.
Another advantage is that the apparatus has to be stable only over individual rotation
periods, in contrast to stationary experiments, which must be stable over 12\,h. Slow (compared to the rotation period) 
systematic variations
of the beat frequency can be eliminated in the data analysis by fitting individual rotation periods. Furthermore, 
in a non-rotating experiment four data values can be obtained from a measurement lasting 12\,h. 

In contrast, a rotating experiment offers the possibility to determine two data points $B(t_i)$ and $C(t_i)$ 
for every rotation ($t_i$ is the mid-time of the rotation period, and the period is chosen much smaller than 12\,h).
Thus, a rotating experiment offers a significant increase in data acquisition rate and thus a reduction 
of statistical noise.
An overall measurement time of 1 year or longer is still necessary for a precise determination of individual coefficients.

\begin{figure}[t]
\epsfxsize=8cm
\epsffile{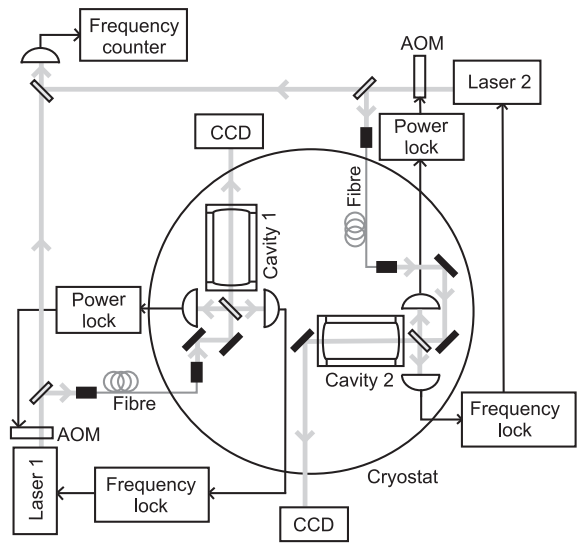}
\caption{\small The experimental set-up. 
Two Nd:YAG lasers are frequency-locked to two sapphire optical cavities located in a cryostat. 
The beams are fed to the resonators via optical fibers.
The cavities' modes can be observed by means of two CCD cameras.
Acousto-optic modulators (AOM) stabilize the power of the beams fed to the resonators.
All components shown are mounted on a rotating table.
\label{set-up:fig}}
\end{figure}

\vskip .2in
\section{Experimental apparatus}
 
Crystalline sapphire optical cavities operated at cryogenic temperature were used.
These cavities 
have previously been used to test LLI with a non-rotating setup \cite{mue03b}
as well as Local Position Invariance \cite{mue02}.
Fig.\,\ref{set-up:fig} shows a schematic of the experiment.
A mechanical rotation stage carries the cryostat (containing the cavities), an associated
optical breadboard and most components (lasers, power supplies, servo systems).
The rotation stage itself rests on an optical table that was not floated.
The cavities were cooled to 3.4~K by means of a pulse-tube cooler
(PTC) \cite{wang97a}, in which He gas is periodically compressed and expanded.
The use of a mechanical cooler has the advantage of continuous operation without
the need to periodically refill cryogens. Furthermore, slow mechanical deformations occurring
in conventional cryostats as a consequence of change in cryogen levels are avoided.

Inside the cryostat vacuum, the two cavities together with various optical components
and photodiodes were mounted on a copper base plate which was supported by rods extending from the top flange of
the cryostat. Copper braids provided thermal contact to the cold head of the cooler,
and offered some compensation from the periodic length modulation
occurring in the cooler due to pressure oscillation in the tubes \cite{lie01}.

The angular range of the stage was limited to about 90 degrees by the He pressure lines connecting the cooler
 to the stationary compressor, but this was sufficient for the purpose of the experiment.
A rotation period 10\,min was chosen for a rotation over $[0^\circ;90^\circ;0^\circ]$. Faster rotation led to disturbances.

The cavities are high-reflection coated for wavelengths around 1$\,\mu$m. In order to avoid rotation-induced variations in the alignment of the laser beams with respect to the cavities that would result in lock errors, 
the laser waves are brought to the cavities through two optical fibers. 
Two diode-pumped monolithic Nd:YAG lasers were locked to the two cavities by means of a Drever-Hall reflection locking scheme \cite{dre83a}.  
The respective photodiodes were located in the cryostat, next to the cavities.
The lasers were frequency-modulated using the piezo actuators glued to the laser crystals as modulators, 
adding the modulation frequencies to the control feedback signals \cite{can95}. 
Parts of the laser beams were superimposed on a fast photodiode producing a heterodyne signal
at the beat frequency $\nu_1 - \nu_2$ between the two lasers
(of the order of 700~MHz). That was measured with a frequency counter.

The periodic pressure modulation (at 1.1$\,$Hz) led to a residual modulation of the beat frequency 
between the two cavity-locked lasers, which amounted to approx.\,200$\,$Hz amplitude. However, this
modulation was later removed using Fourier filtering.
The typical Allan standard deviation of the beat frequency (for nonrotating resonators) was 2.5~Hz 
for an integration time of 30\,s, and 5\,Hz ($\simeq2\cdot10^{-14}$) for an integration time of 300\,s, 
(half of the rotation period).

The dependence of the cavity resonance frequencies on the beam powers was measured to be $\sim
 8\,$Hz per $\mu$W impinging on the cavities.
The laser powers  were  actively stabilized using acousto-optical-modulators (AOM),
also serving as optical isolators.  
At a typical working power of 50-100$\,\mu$W onto the cavities, the influence of residual power changes was thereby reduced 
to the level of 0.1~Hz ($0.4\cdot10^{-15}$) for integration times of hundreds of seconds.

The dependence of the beat frequency on the temperature of the baseplate supporting the cavities was on the order of 1.5~Hz/mK.

Active stabilization resulted in an instability of 45~$\mu$K at 300\,s integration time, and thus a temperature-induced
beat frequency instability on the order of 0.07~Hz ($2\cdot10^{-16}$). 

The sensitivity of the beat frequency to tilt of the table supporting the apparatus was \,0.06\,Hz/$\mu$rad ($2\cdot10^{-16}/\mu$rad). 
The tilts along two horizontal axes were monitored continuously during the rotations for later decorrelation. 
Typical tilt modulation during a rotation was 50 $\mu$rad peak-to-peak, corresponding to a inferred beat modulation of several Hz. 
However, for the data analysis only the beat frequency modulation components at $2\theta$ are relevant; for these the 
inferred peak-to-peak amplitudes were typically less than 1\,Hz.

\begin{figure}[t]
\ \hskip-1.5cm\epsfxsize=11cm\epsffile{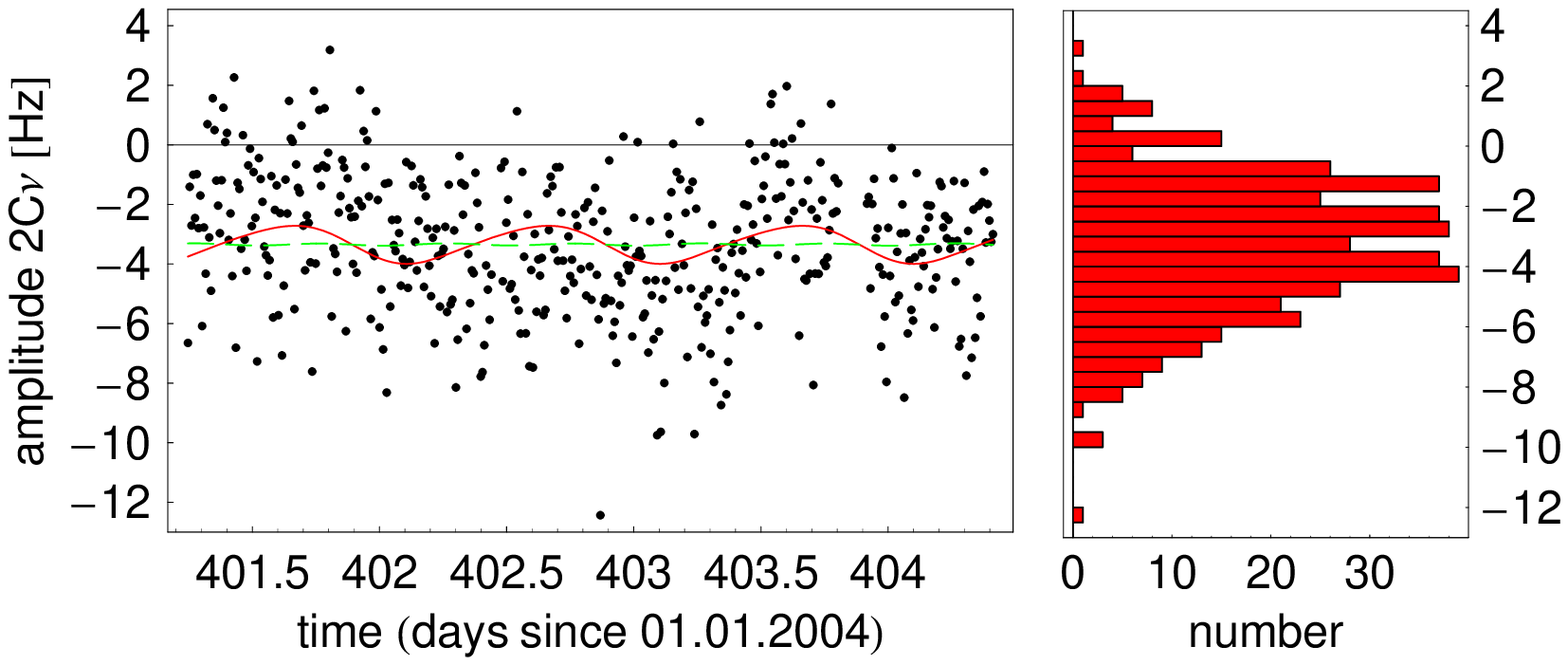}

\ \hskip-1.5cm\epsfxsize=11cm\epsffile{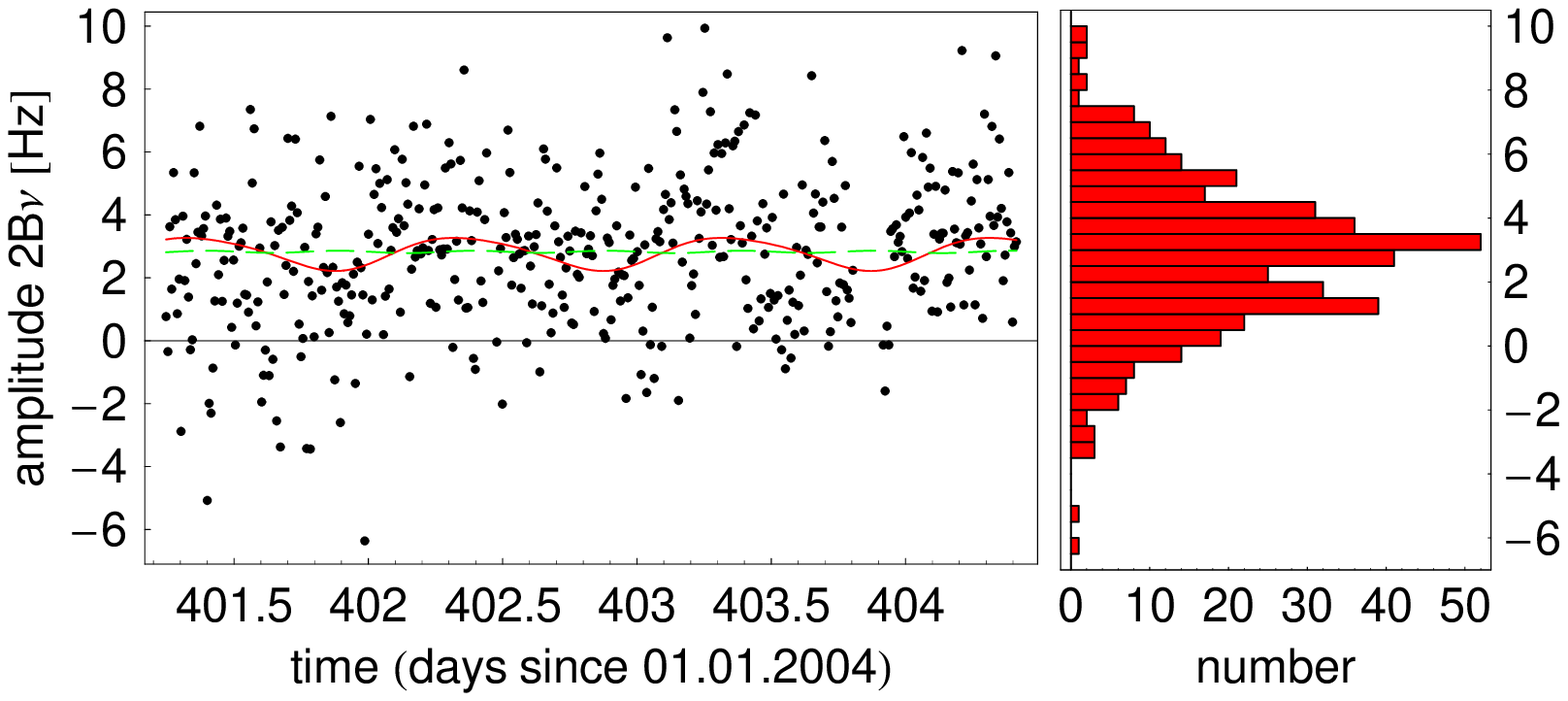}
\caption{\small Measured $2B\nu$ and $2C\nu$ amplitudes of spatial anisotropy for individual rotations, and corresponding histograms. 
The fit error for each data point is less than 1\,Hz. Full line: SME model plus systematic effect.
Dashed line: RMS model plus systematic effect. 
Mean values: $\langle 2B\rangle$= 2.8 Hz,
$\langle 2C\rangle$= -3.3\,Hz.}
\label{2B.fig}
\end{figure}

\vskip .2in
\section{Results}

The data analyzed in this publication was collected during very stable operation 
from 4 February 2005 to 8 February 2005 for 76 hours. 
Frequency sampling was at 1 s intervals. The measured tilts were time-averaged, 
weighted with the measured beat frequency sensitivities and subtracted from the beat data.
Then, the beat modulation at 1.1 Hz, caused by the PTC, was removed. After removal of clearly disturbed rotations,
the remaining 432 rotation periods $\theta=[0^\circ;90^\circ;0^\circ]$ (labelled by $i$) were each
least-squares fitted with functions $a_i t + 2 B(t_i) \sin 2\theta(t)+2 C(t_i) \cos 2\theta(t)$, 
where the coefficients $a_i$ quantifiy a (slowly varying) linear drift.

The obtained amplitude sets $\{2B(t_i)\}$, $ \{2C(t_i)\}$ are shown in Fig.\,\ref{2B.fig}.
The data were analysed in the following ways.

(i) Within the RMS-framework, for two orthogonal cavities located on Earth and considering the preferred frame mentioned above,
the beat frequency can be written as in Eq.(\ref{eq1}) with
\begin{displaymath}
2C(t)=(1/2-\beta+\delta)({v^2}/{c^{2}_0})
(\gamma_0+\gamma_1 \cos \omega_{\oplus}T_{\oplus}+
\end{displaymath}
\begin{displaymath}
\gamma_2 \cos 2\omega_{\oplus}T_{\oplus}+\sigma_1 \sin \omega_{\oplus}T_{\oplus}+
      \sigma_2 \sin 2\omega_{\oplus}T_{\oplus}),
\end{displaymath}
\begin{equation}
2B(t)=(1/2-\beta+\delta)({v^2}/{c^{2}_0})
(\gamma_3 \cos \omega_{\oplus}T_{\oplus}+ 
\end{equation}
\begin{displaymath}
\gamma_4 \cos 2\omega_{\oplus}T_{\oplus}+\sigma_3 \sin \omega_{\oplus}T_{\oplus}
+\sigma_4 \sin 2\omega_{\oplus}T_{\oplus})\ ,
\end{displaymath}

\noindent where 
$\gamma_0=\sin^2 \chi (3 \cos^2 \Theta -1)/4$, 
$\gamma_1=(1/2)\cos \Phi \sin 2\Theta \sin 2\chi$, 
$\gamma_2=\cos 2\Phi \cos^2\Theta(\cos 2\chi -3)/4$, 
$\gamma_3=\sigma_3\tan\Phi $,
$\gamma_4=-\sigma_4 \tan 2\Phi$, 
$\sigma_1=\gamma_1\tan \Phi $, 
$\sigma_2=\gamma_2\tan 2\Phi$,
$\sigma_3=\cos\Phi \sin\chi \sin 2\Theta$, 
$\sigma_4=\cos^2 \Theta \cos\chi \cos 2\Phi$\ . 

\noindent  $T_{\oplus}$ is the time since the beginning of the data plus an offset that accounts for a time difference since the coincidence 
of the lab's $y$ axis with the $\hat{Y}$ axis of the Sun-centered system \cite{Kostelecky:2002hh}. The direction of  the Sun's velocity $\vec v$ 
 relative to the cosmic microwave background is given by 
the right ascension $\Phi=168^\circ$ and the declination $\Theta=-6^\circ$. 
In fitting the above expression to our data, we take into account that the finite average values of $2B$ and $2C$ are likely to be
due to a (constant) systematic error. We model this by adding a contribution $b_{syst}\sin 2\theta+c_{syst}\cos 2\theta$ 
to Eq.(\ref{eq1}), and fit $b_{syst},\,c_{syst},\,\beta - \delta-\frac{1}{2}$. 
Effectively, this means that only the modulation of the $\{2B,2C\}$ amplitudes by Earth's rotation 
contributes to the fit result for $\beta - \delta-\frac{1}{2}$.  

\begin{table}[b]
\caption{\label{tab:table1}
\small Fourier amplitudes determined from the experiment. 
All quantities are in units of $10^{-16}$.}
\begin{tabular}{l|l|r|r|r}
 & Basis & Value& Statistical\,error & Systematic\,error\\
\hline
$C_0$ & 1 			                   &$-59$ & $3.4$ & $3$  \\
$C_1$ & $\sin{(\omega_{\oplus}T_{\oplus})}$ & $-3$ & $1.5$  & $0.5$  \\
$C_2$ & $\cos{(\omega_{\oplus}T_{\oplus})}$ & $11$ & $2$  & $0.5$ \\
$C_3$ & $\sin{(2\omega_{\oplus}T_{\oplus})}$ & $1$ & $2$  & $0.5$  \\
$C_4$ & $\cos{(2\omega_{\oplus}T_{\oplus})}$ & $0.1$ & $2$ & $0.2$
\end{tabular}
\end{table}

The fit  yields $\beta - \delta-\frac{1}{2}=(+0.5\pm 3)\cdot10^{-10}$ and is shown in Fig.\,\ref{2B.fig}. Due to the experimental error in the determination of the 
tilt sensitivities there is an additional (systematic) error of $\pm 0.7\cdot10^{-10}$.
Our result is about factor 10
 lower than the previous best results $(-1.2\pm1.9\pm1.2)\cdot10^{-9}$ \cite{Wolf04b}  and $(-2.2\pm1.5)\cdot10^{-9}$ 
\cite{mue03b}.

(ii) Using all our data we can obtain a fit of the 5 Fourier amplitudes of 
$2C(t^*)$ at the time $t^*$ $\simeq$  6 February 2005,  see Table \ref{tab:table1}. 
These amplitudes are linear combinations of the $\tilde{\kappa}_{e-}$ and 
$\tilde{\kappa}_{o+}$ coefficients, where the weights of the latter depend on Earth's orbital phase
 (see Appendix E of \cite{Kostelecky:2002hh}). 
For the fit we make use of the approximate relationship between the functions $B(t)$ and $C(t)$ \cite{foot2} and we
include a coefficient $b_0$ in $B(t)$ in order to describe a systematic effect. 
Note that due to the limited extent of the data over time, 
the frequencies $\omega_{\oplus}+\Omega_{\oplus}$ and $\omega_{\oplus}-\Omega_{\oplus}$ cannot be 
distinguished and therefore it is not possible to extract the individual coefficients $\tilde{\kappa}_{e-}$
 and $\tilde{\kappa}_{o+}$ from the $C_i$ coefficients.

(iii) The results of the microwave cryogenic experiment \cite{Wolf04} found the elements of
$(\tilde{\kappa}_{e-})$ (except for $(\tilde{\kappa}_{e-})^{ZZ})$ and the elements of $\beta_\oplus$ ($\tilde{\kappa}_{o+}$)
to be zero within several parts in $10^{-15}$. 
If we assume these elements to be zero, we can use only the $C_0$-coefficient to determine
$(\tilde{\kappa}_{e-})^{ZZ}$, {\it i.e.} Eq.(\ref{eq2}) is truncated at the first term.
The result is $(\tilde{\kappa}_{e-})^{ZZ} = (-2.0\pm0.2)\cdot 10^{-14}$. 
As this nonzero value is likely due to a systematic effect, we may state that the value of 
$|{(\tilde{\kappa}_{e-})^{ZZ}}|$ is probably less than $2\cdot 10^{-14}$.

We believe the nonzero averages of the $2B$ and $2C$ amplitudes are due to a systematic effect of thermal origin. 
A (small) thermal gradient is present in the laboratory, through which the apparatus rotates. 
The magnitude of the effect is consistent with our measured temperature 
sensitivity and the temperature modulation amplitude measured at the top of the apparatus.

Overall, our experiment is 
limited firstly by the slow maximum rotation speed allowed by the system. A faster, variable rotation speed 
would in principle allow a more detailed study of the systematic effects or a partial suppression thereof. 
However, faster rotation could not be used with our setup.
Secondly, 
the laser frequency lock stability is limited by the small signal-to-noise ratio (low cavity throughput), which limits the ability
to more precisely characterize the systematic effects. 

In conclusion, we have described a Michelson-Morley-type experiment that has led to the first determination of
a limit for $(\tilde{\kappa}_{e-})^{ZZ}$ of the Standard Model Extension, $|\tilde{\kappa}_{e-}^{ZZ}| < 2 \cdot 10^{-14}$. 
An analysis of the 
data in the RMS framework yielded an upper limit for a hypothetical direction dependence of $c$
 an order of magnitude lower than the best previous measurements. 
We note that the present results were achieved using a measurement 
duration one hundred times shorter. 
The accuracy of the experiment appears to be limited by a thermal systematic effect. 
A significant improvement of our results would require major modifications to the apparatus.

\vskip .2in
\section{Acknowledgments}
 {\small We thank L. Haiberger for contributions to the cryostat development, A. Nevsky and C. L\"ammerzahl for discussions, 
and G. Thummes for his helpful assistance.
P.A. was supported by a DAAD fellowship, M.O. by a Heinrich-Hertz Foundation fellowship. 
This research was part of the Gerhard-Hess Program of the German Science Foundation.}


\begin{thebibliography}{10}
\bibitem{foot2}
The Fourier amplitudes of the $B$ data can be obtained from the $C$ amplitudes if the contributions 
due to the velocity of the laboratory with respect to the Earth's center, $\beta_L\approx 10^{-6}$, are neglected. 
Then 
$B_0=0$,
$B_1 = -C_2/\cos{\chi}$,
$B_2 = C_1/\cos{\chi}$, 
$B_3 = -2C_4\cos{\chi}/(1+\cos^2{\chi})$ and
$B_4 = 2C_3\cos{\chi}/(1+\cos^2{\chi})$.
\end{thebibliography}
\end{document}